\documentclass[a4paper,fleqn]{cas-dc}

\usepackage[authoryear]{natbib}
\usepackage{xcolor}

\def\tsc#1{\csdef{#1}{\textsc{\lowercase{#1}}\xspace}}
\tsc{WGM}
\tsc{QE}
\usepackage{tablefootnote}
\let\oldAA\AA
\renewcommand{\AA}{\text{\normalfont\oldAA}}

\begin{document}
\let\WriteBookmarks\relax
\def\floatpagepagefraction{1}
\def\textpagefraction{.001}

\shorttitle{Electrochemical Measurement of the Electronic Structure of Graphene}    

\shortauthors{Paulo R. Bueno}  

\title [mode = title]{Electrochemical Measurement of the Electronic Structure of Graphene via Quantum Mechanical Rate Spectroscopy}

\author[]{Laís Cristine Lopes}
\author[]{Edgar Pinzón}
\author[]{Gabriela Dias-da-Silva}
\author[]{Gustavo Troiano Feliciano}
\author[]{Paulo Roberto Bueno}
\ead{paulo-roberto.bueno@unesp.br}
                      
\cormark[1]

\ead{paulo-roberto.bueno@unesp.br}

\affiliation[1]{organization={Department of Engineering, Physics and Mathematics, Institute of Chemistry, São Paulo State University}, 
            city={Araraquara},
%          citysep={}, % Uncomment if no comma needed between city and postcode
            postcode={14800-060}, 
            state={São Paulo},
            country={Brazil}}


\begin{abstract}[summary]
Quantum-rate theory defines a quantum mechanical rate $\nu$ that complies with the Planck--Einstein relationship $E = h\nu$, where $\nu = e^2/hC_q$ is a frequency associated with the quantum capacitance $C_q$, and $E = e^2/C_q$ is the energy associated with $\nu$. Previously, this definition of $\nu$ was successfully employed to define a quantum mechanical meaning for the electron-transfer (ET) rate constant of redox reactions, wherein faradaic electric currents involved with ET reactions were demonstrated to be governed by relativistic quantum electrodynamics at room temperature~\citep{Bueno-2023-3}.

This study demonstrated that the definition of $\nu$ entails the relativistic quantum electrodynamics phenomena intrinsically related to the perturbation of the density-of-states $\left( dn/dE \right) = C_q/e^2$ by an external harmonic oscillatory potential energy variation. On this basis, the electronic structure of graphene embedded in an electrolyte environment was computed. The electronic structure measured using quantum-rate spectroscopy (QRS) is in good agreement with that measured through angle-resolved photo-emission spectroscopy (ARPES) or calculated via computational density-functional theory (DFT) methods.

Electrochemical QRS has evident experimental advantages over ARPES. For instance, QRS enables obtaining the electronic structure of graphene at room temperature and in an electrolyte environment, whereas ARPES requires low temperature and ultrahigh-vacuum conditions. Furthermore, QRS can operate \textit{in-situ} using a hand-held, inexpensive piece of equipment, whereas ARPES necessarily requires expensive and cumbersome apparatus.

\end{abstract}


%\item X.

%Keywords

\maketitle

% Main text
\section{Introduction}\label{sec:introduction}

This section provides a straightforward introduction of the quantum-rate concept~\citep{Bueno-book-2018, Bueno-2020}. Its applications in the field of electrochemical reactions are detailed in the literature~\citep{Bueno-2023-3}, which provides insights as to how this rate concept is indicative of the electrodynamics of electron-transfer reactions. Furthermore, a relevant prior study focused on the application of the quantum-rate theory and a $\nu$ quantum-rate concept to graphene~\citep{Bueno-2022}. This study focused on the application of this concept as a tool for the \textit{in-situ} measurement of the electronic structure of graphene. We are referring to this tool as quantum-rate spectroscopy (QRS).

\subsection{Quantum Mechanical Rate}\label{sec:QRconcept}

Quantum-rate theory~\citep{Bueno-2023-1, Bueno-2023-2} is based on the first-principles concept of the quantum mechanical rate, which is defined as the ratio between the reciprocal of the von Klitzing constant $h/e^2 \sim$ 25.8 k$\Omega$ and quantum capacitance.

\begin{equation}
\label{eq:nu}
	\nu = \frac{e^2}{h C_q} = \frac{E}{h},
\end{equation}

\noindent where $e$ is the elementary charge of the electron, $h$ is Planck’s constant, $C_q$ is the quantum capacitance, and $E = e^2/C_q$ is the energy state of a single electron in a single quantum channel with Fermi velocity $c_*$. Thus, this $\nu$ concept helps establish a direct correlation with the relativistic electron transport occurring between the intra- and intermolecular quantum states. 

Notably, this type of electrodynamics is different from the classical transport of electrons (where the drift velocity of the carries is of $\sim$ 10$^{-3}$ m/s) because it predicts transmittance of electrons between quantum states connected via quantum channels where the velocity of the carriers is the Fermi velocity, i.e. $\sim$ 10$^6$ m/s. In other words, the velocity of the charge carriers, in addition to being coherent with a time-dependent potential source of perturbation, follows a Fermionic velocity that is several others of the magnitude of that predicted by the classical drift velocity. In other words, the quantum electrodynamics predicts that the quantum rate theoretical concept is not related to the traditional drift velocity of electrons (which is several orders of magnitude lower than the Fermi velocity), for instance, obtained in the electric current flowing in metallic structures. Therefore, the Fermi velocity of transport of charge carriers within molecular structures involves relativistic electrodynamics, wherein an external electric time-dependent potential perturbation is the origin of a coherent transport such that the flowing of charge between quantum states is in phase coherence with the electronic structure, which thus can be determined by measuring $C_q$ at a particular low-frequency equilibrium perturbation $\omega_0$, as will be discussed in subsequent sections.

\subsection{Relativistic Electrodynamics}\label{sec:Relectrodynamics}

The relativistic electrodynamics within Eq.~\ref{eq:nu}, that is, $E = h\nu = e^2/C_q$, denotes the relationship between energy $e^2/C_q$ of the electronic structure and a frequency $\nu$ with which it can be perturbed. This, combined with the De Broglie relationship $\textbf{p} = \hbar \textbf{k}$ of particle--wave duality, entails a relativistic quantum electrodynamics of charged particles within the electronic structure -- that can be quantified through its quantum capacitance $C_q$ -- owing to the linear proportionality between the energy and wave vector \textbf{k}, i.e.

\begin{equation}
 \label{eq:Planck-Einstein}
	E = \hbar \textbf{c}_* \cdot \textbf{k} = \textbf{p} \cdot \textbf{c}_*,
\end{equation}

\noindent where $\hbar$ = $h/2\pi$. The intrinsic relativistic nature of Eq.~\ref{eq:Planck-Einstein} implies that electron spin degeneracy must be considered in the electrodynamics. 

As described in the next subsection, consideration of the spin degeneracy $g_s$ combined with $C_q$ facilitates the definition of a concept of $\nu$ for describing the relativistic quantum electrodynamics. Owing to a phase coherence between the relativistic quantum transport of charge and $C_q$, the information on the electronic structure can be obtained.

\subsection{Spin Energy State Degeneracy and Conductance Quantum}\label{sec:RC}

To consider electron spin degeneracy $g_s$ in the definition of $\nu$, we first express the energy of the electronic structure as $E = g_s e^2/C_q$; hence, Eq.~\ref{eq:nu} becomes

\begin{equation}
\label{eq:nu-G}
	\nu = \frac{g_s e^2}{h C_q} = \frac{G_0}{C_q},
\end{equation}

\noindent where $G_0 = g_s e^2/h$ is identified as the conductance quantum, which is a constant with the magnitude of $\sim$77.5 $\mu$S. Further, $R_q = 1/G_0 = h /g_s e^2$ is a resistive quantum limit with the magnitude $\sim$12.9 k$\Omega$. 

In summary, the introduction of the spin degeneracy implies a characteristic quantum resistive-capacitive (RC) time constant $\tau = R_q C_q = 1/\nu$ that depends only on the meaning of $C_q$ to be resolved. The relationship between $R_q$ and $C_q$ is phase coherent and can be experimentally resolved by electric time-dependent perturbation methods. As shown in subsequent sections, the analysis of the quantum RC circuit for graphene helps reveal the electronic structure and thus the energy and wave vector relationship associated with particle dynamics. This facilitates the measurement of the electronic structure by measuring the quantum characteristic $\tau$ of the system.

\subsection{A Three-dimensional Analysis of Graphene Wave Vector}\label{sec:3D-graphene}

In a previous work~\citep{Bueno-2022}, the meaning of the quantum rate concept within 2D pristine graphene electronic structure model was demonstrated. Graphene was specifically selected so that a three-dimensional interpretation of \textbf{k} could be considered. A third dimension for $\textbf{k}$ was introduced to account for the influence of the thickness of graphene on its electronic structure. This is particularly important if measurements are conducted in an electrolyte environment wherein the screening of the electric field within the 2D structure of graphene would be conducted through the electrolyte. Therefore, the presence of an electrolyte is key to studying graphene in its pristine form; a significantly small (energy) perturbation needs to be imposed to compute the quantum properties of graphene. 

Accordingly, the thickness must be considered to decouple the space and time dependence of the bispinors; this helps resolve the Dirac equation, considering $\varphi_e = e^{-j\omega_e t}\psi_e$ and $\varphi_h = e^{j\omega_h t}\psi_h$, where an energy of $\hbar\omega_e = E_e - eV_{ch} + m_c c_*^2$ was considered for electrons ($e$), and $\hbar\omega_h = - E_h + eV_{ch} + m_c c_*^2$ was considered for hole ($h$) carriers~\citep{Bueno-2022}. Supposedly, the time perturbation of the external potential $V(t)$ is markedly lower than $E_e/e$ and $E_h/e$; this is a good approximation for interpreting the impedance-derived capacitance spectra for graphene, wherein the requirement of linearity between the perturbation and response signals is fundamental for the method to be valid. In other words, small-signal-amplitude perturbation of the system is a key requirement. This was the only constraint imposed to solve the Dirac equation wherein both the amplitude- and frequency-equilibrium perturbations are small. The equilibrium frequency perturbation for acquiring the density-of-states (DOS) of the system, for instance, is at least three times lower than $k_B T$. Therefore, the solution to the Dirac equation for the aforedescribed boundary condition, considering the influence of the thickness and a small perturbation for acquiring system information, was theoretically converted into the Helmholtz equation such that

\begin{equation}
    \nabla^2 \psi + |\textbf{k}|^2 \psi = 0,
    \label{eq:Helmholtz}
\end{equation}

\noindent where the wave vector $|\textbf{k}|$ is resolved as a three-dimensional space variable separated from time such that 

\begin{equation}
    |\textbf{k}| = \left( \frac{E_eE_h}{c_*^2 \hbar^2} \right)^{1/2},
    \label{eq:k}
\end{equation}

\noindent encompassing both the relativistic and non-relativistic cases of the energy and wave vector dispersion relationships~\citep{Bueno-2022}. For instance, if we consider the symmetry of electrons and holes in graphene with respect to the Dirac point, $E_e = E_h = E$, and the dispersion relationship in Eq.~\ref{eq:k} conforms to $E = \hbar c_* |\textbf{k}|$. This is in agreement with that predicted using the quantum rate theory, as discussed in the introduction section, where $E = \hbar c_* |\textbf{k}| = h \nu$, with $\nu = 1/\tau$ as the meaning of the quantum rate.

The most important result arising from a three-dimensional interpretation of $\textbf{k}$ in a pristine single-layer graphene is that the effect of an external potential $V(t)$ perturbation in the spinor $\psi$ can be considered such as that the effect of a $V(t)$ stimulus in the plane $x-y$ can be measured through a time perturbation in the $z$ direction. This is because, mathematically, the plane $x-y$ can be modelled separately in the perpendicular $z$ direction. The stimulus in the plane $x-y$ can be observed by invoking symmetry properties of the spinor $\psi$, and the quantum rate can be spatially interpreted within the spatial quantum electrodynamics setting of Eq. ~\ref{eq:Helmholtz}. 

For instance, consider $\psi(x, y, z) = \phi(x,y)\zeta(z)$ such that $\zeta(z) = \zeta_0 \sin(k_z z)$ and $\phi(x,y) = \phi_0 \sin(|\textbf{k}|_x x + |\textbf{k}|_y y)$, where $|\textbf{k}|_x^2 + |\textbf{k}|_y^2 + |\textbf{k}|_z^2 = |\textbf{k}|^2 = E_eE_h / (c_*\hbar)^2$. This interpretation enables the imposition of boundary conditions for a 2D single-layer graphene (SLG), allowing for the use of a quantization mode that obeys $|\textbf{k}|_z = 2\pi n_{\nu}/ L_z$, where $L_z$ is the thickness of the SLG (in the z-direction, perpendicular to the $x-y$ plane) and where $n_{\nu}$ is a non-zero natural number associated with the quantum number (or quantum channel) modes for charge motion in the $z-$direction perpendicular to the $x-y$ plane of graphene. In this $z-$direction, there is a time $t_z$ related to the characteristic relaxation time for the transport of charge carriers within the thickness $L_z$ such that $L_z = c_* t_z$ with $\tau_z = t_z / 2\pi$ and $|\textbf{k}|_z =  n_{\nu} / c_* \tau_z$. 

Notably, charge perturbation transversal to the $x-y$ plane of conductance in graphene, according to $\zeta(z)$, describes a stationary harmonic motion perturbation that propagates as a wave in the $x-y$ plane of graphene at a velocity $c_*$, promoting the charge flowing as a displacement current in the z-direction, which is coupled to the $x-y$ plane. In other words, the boundary condition that establishes a highly slow oscillatory time perturbation with a corresponding energy excitation at least three times lower than $k_B T$ (referred to as the equilibrium frequency of perturbation) imposes that the fundamental stationary quantum number state $n$ of the $x-y$ plane is the same of that of the z-direction, such as that $n = n_{\nu}$. 

Therefore, the correlation between the quantum number states of the $x-y$ plane and those of the $z$ direction implies that the induced displacement electric current established by an external perturbation in the thickness propagates with the same velocity $c_*$ in the plane. This oscillatory electric potential perturbation in the z-direction facilitates the measurement of the response properties of DOS in the $x-y$ plane of graphene.

\subsection{Quantum Capacitance in a Relativistic Situation}\label{sec:Cq}

A quantum capacitance $C_q$ can be associated with relativistic dynamics by noting that $\nu = \left( E/h \right) = \left( c_*/L_z \right)$ can be directly correlated with the $L_z$ length of graphene and $E = e^2/C_q$. As discussed in the previous section, if the fundamental pristine quantum states of graphene are perturbed, the quantum numbers are the same or at least proportional such that the perturbation in the thickness propagates in the plane. The number of quantum channels associated with the electronic structure of graphene in the plane permits the relativistic transport of charge carriers, including electrons and holes, owing to $\nu = \left( c_*/L_z \right) = e^2/hC_q$, which directly yields the energy of the channel when multiplied by the Planck’s constant; hence, it is dependent on $C_q$. The aforedescribed analysis implies that $L_z = n \lambda$, where $n$ is the number of quantum states within $L_z$ and $\lambda$ is the wavelength directly related to $|\textbf{k}| = 2\pi/\lambda$, where \textbf{|\textbf{k}|} is the modulus of the wave vector \textbf{k}.

Accordingly, the energy can be written as a function of the number of states (or quantum channels) $n = L_z/\lambda$ such as $E = n h \left( c_* /\lambda \right) = n \left( h\nu \right)$, where $E = h \left( c_*/\lambda \right) = h \nu$ is the energy $e^2/C_q$ of a single state. Hence, the DOS can be obtained simply as $\left( dn/dE \right) = C_q/e^2$, and the quantum capacitance is thus defined as

\begin{equation}
\label{eq:Cq}
	C_q = e^2 \left( \frac{dn}{dE} \right),
\end{equation}

\noindent from where $C_q$ is directly proportional to the DOS. 

The next section describes the DOS and quantum capacitance of graphene, as well as the characteristic time constant associated with a coherent transport of carriers with the electronic structure defined by $C_q$.

\begin{figure}[h]
\centering
\includegraphics[width=8cm]{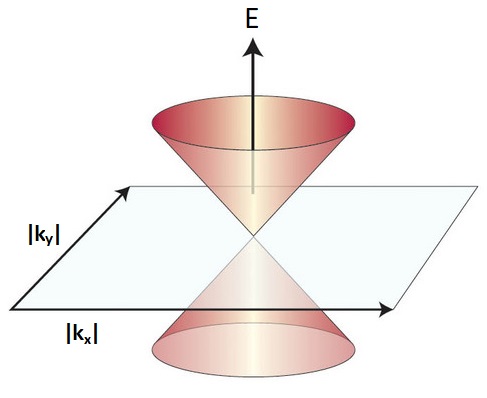}
\caption{Representation of the relativistic quantum electrodynamics $E(\textbf{k}) = \hbar \textbf{c}_* \cdot \textbf{k}$ of graphene, which ignores the thickness contribution. Here, for the $x-y$ plane of graphene, the modulus of $\textbf{|k|}_{x-y}$ is given by $\left( |\textbf{k}_x| + |\textbf{k}_y| \right)^{1/2}$. In the quantum-rate theory of graphene, the conductance in the $x$ and $y$ directions are quantized by the integer number of the conductance quantum $G_0$ as well as the rate $\nu$.}
\label{fig:E-k}
\end{figure}

\subsection{Two-dimensional Density-of-States of Graphene}\label{sec:DOS-graphene}

\begin{figure}[h]
\centering
\includegraphics[width=8cm]{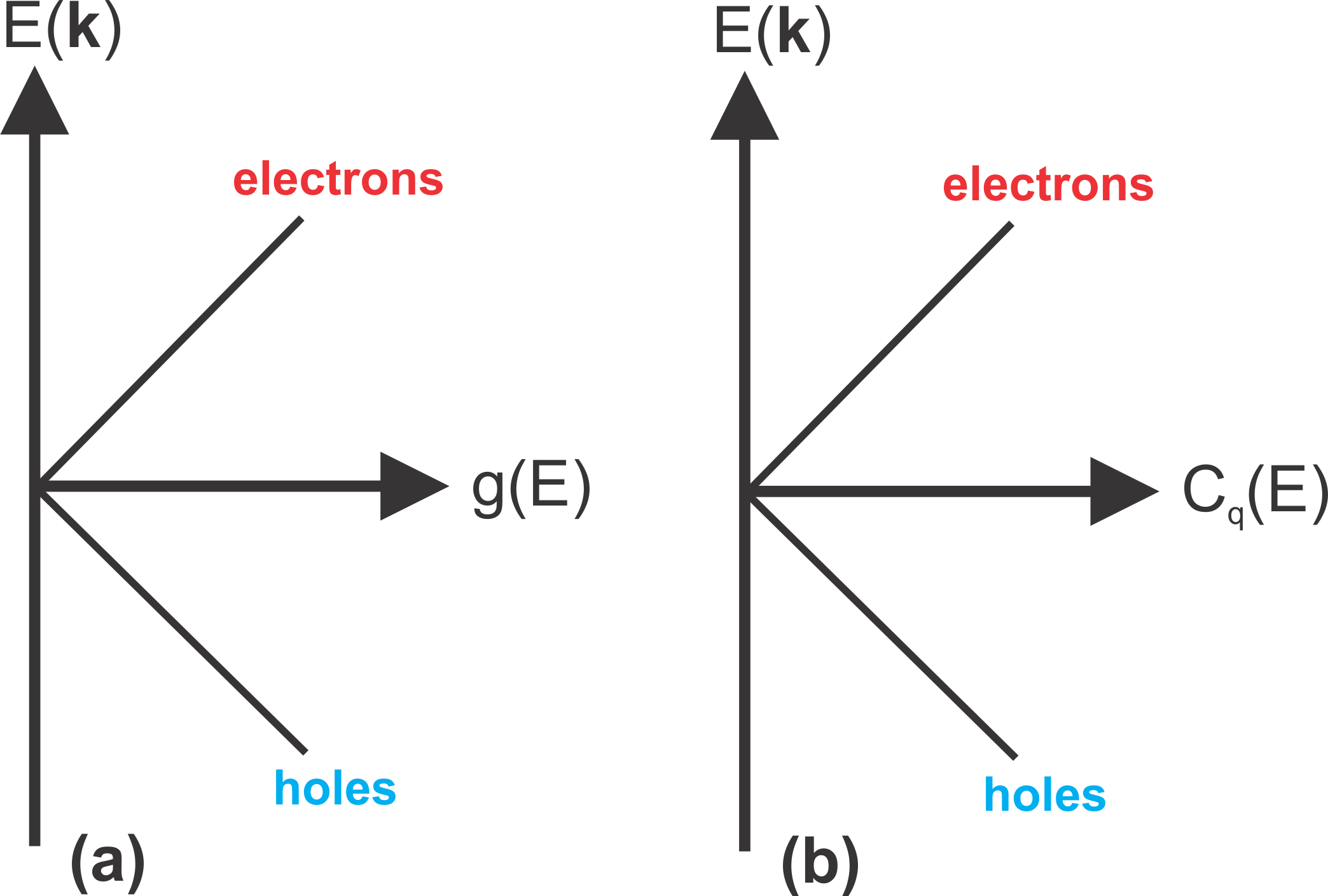}
\caption{Representation of the well-known V-shape of graphene, wherein an experimental measurement of electron or hole injections into the electronic structure allows access to the DOS function $g(E)$. (a) shows the traditional analysis of $E(\textbf{k})$ \textit{versus} $g(E)$, in which $\delta E(\textbf{k}) = e \delta V$ corresponds to different charge state and energy $\Delta E(\textbf{k})$ occupancies of the electronic structure of graphene as a function of the bias potential $\Delta V$ concerning the Dirac point used as zero of energy reference. (b) Depicts the analysis proposed by the quantum-rate theory where the quantum capacitance $C_q$ is measured in different energy levels of the occupancy of the electronic structure by electron or holes in an equilibrium charge occupancy situation driven by a different bias energy $e \delta V$ established with respect to the Dirac point as the energy reference for the vacuum level. As $C_q(E)$ directly correlates with a coherent $\nu = 1/R_qC_q$ perturbation of the electronic structure, the measurement of $C_q$ enables obtaining information of the momentum through Eq.~\ref{eq:tau-q(k)}; thus, spatial information can be obtained.}
\label{fig:E(k)-g(E)}
\end{figure}

The relativistic quantum dynamics of Eq.~\ref{eq:Planck-Einstein} defines $E/|\textbf{k}|_{x-y} = dE/d|\textbf{k}|_{x-y} = \hbar c_*$ in the $x-y$ plane of graphene, as depicted in Figure~\ref{fig:E-k}, where $|\textbf{k}|_{x-y} = \left( |\textbf{k}|_x + |\textbf{k}|_y \right)^{1/2}$ is the modulus of the resultant wave vector. The number of states $n$(|\textbf{k}|$_{x-y}$)$d$|\textbf{k}|$_{x-y}$, accounted for in the $k-$space, per differential interval of d|\textbf{k}|$_{x-y}$ in the $x-y$ plane of graphene, can be expressed as~\citep{Castro-Neto-2009}

\begin{equation}
\label{eq:n(k)}
	n\left( |\textbf{k}|_{x-y} \right)d|\textbf{k}|_{x-y} = g_s g_v \frac{A}{2\pi} \left[ \frac{E dE}{\left( \hbar c_* \right)^2} \right],
\end{equation}

\noindent where $A = L_x L_y$ is the area of the $x-y$ plane, wherein $L_x$ and $L_y$ are the quantum channel lengths in the $x$ and $y$ directions of the plane, respectively. $g_v$ is the degeneracy of the valley and is related to the conical symmetry above and below the $x-y$ plane. There are different interpretations for the consideration of $g_v$. The most accepted interpretation is that $g_v$ implies an ambipolar electric current due to electron and hole charge carriers that are directly associated with the conical symmetry of $K$ point space, as depicted in Figure~\ref{fig:E(k)-g(E)}. 

This conical symmetry~\citep{Bostwick-2007, Geim-2010}, related to electrons and holes in graphene, indicates the well-known semimetal characteristics of conduction and valence bands with a singularity at the Dirac point. The spatial geometric interpretation implies six locations in the momentum space, corresponding to the vertices of the hexagonal Brillouin zone~\citep{Bostwick-2007, Geim-2010}. These six locations can be divided into two non-equivalent sets of three points, resulting in two sets labeled $K$ and $K'$, that is, a valley $g_v$ degeneracy.

Within the quantum rate theory, the $g_v$ degeneracy has an equivalent interpretation to that of $g_e$ or $g_r$ degeneracy attributed to the ambipolar characteristics of electric current associated with redox reactions, which hence is also based on the existence of resonant electric currents, i.e., of time-dependent (displacement) electric currents within the electrochemical junction. This type of resonant electric current is owing to the existence of two charge carriers (electrons and holes) that promote a net current (denoted here as $i_0$) in the interface within a redox reaction. In any of the cases (quantum electrodynamics of graphene~\citep{Bueno-2022} and electrochemical reactions~\citep{Bueno-2023-3}), the origin of this displacement electric current is owing to the role played by the electrolyte that allows the superimposition of the electrostatic $C_e$ and quantum capacitive $C_q$ modes of charging the molecular junction states. In this situation, the elementary charge $e$ is subjected to an equivalent electric potential, i.e., $e/C_e \sim e/C_q$, conducting locally to $C_e \sim C_q$. 

Hence, the equivalent $C_\mu$ capacitance of the interface is $1/C_\mu = 2/C_q$, with an energy degeneracy of $e^2/C_\mu = 2e^2/C_q$, where 2 is accounted as $g_e$ (which is equivalent ot $g_v$ in the case of graphene) in the formulation of $\nu = g_e G_0/C_q$ concept. This energy degeneracy is equivalent to the previous electrical current degeneracy consideration of the origin of $g_e$ (for electrochemical reactions occurring in redox-active monolayer junctions~\citep{Bueno-2022}) or $g_v$ (for graphene junction~\citep{Bueno-2023-3} embedded in an electrolyte environment) because $1/C_\mu = 2/C_q$ implies $C_q = 2C_{\mu}$. Noting that the capacitance of the interface is a result of the equivalent contribution of $C_e$ and $C_q$, there is an electric current degeneracy for charging the quantum capacitive states of the interface that is proportional to $C_\mu$ with an electric current of $i_0 = C_{\mu}s = (1/2)(C_q)s$, which is equivalent to $2 i_0 = C_q s$, where $s = dV/dt$ is the time potential perturbation imposed to the interface to investigate the energy $E = e^2/C_q$ level\footnote{Note that $E$ has an intrinsic degeneracy, i.e. $g_s g_v E$ for the case of graphene~\citep{Bueno-2022} whereas it is $g_s g_e E$ for the case of redox reactions~\citep{Bueno-2023-3}. Accordingly, the quantum rate theory accommodates both quantum electrodynamics measured in electrochemical reactions and graphene within an equivalent energy degeneracy.}, electronically coupled to the electrode states. Note that it is owing to the equivalence of $e^2/C_e$ and $e^2/C_q$ energy states that there is an energy degeneracy that inherently permits two electric currents contributing to the net (electrons and holes) current $i_0$ of the interface, which is thus said to be an ambipolar electric current.

Notably, as indicated by Eq.~\ref{eq:n(k)}, $\left[ n \left( |\textbf{k}|_{x-y} \right)d|\textbf{k}|_{x-y} \right] /dE$ is the only function of energy. 

\begin{equation}
\label{eq:g(E)}
	g(E) = g_s g_v \left( \frac{A}{2\pi} \right) \left[ \frac{E}{(\hbar c_*)^2} \right],
\end{equation}

\noindent which is known as the DOS. Particularly, $g(E)$, in Eq.~\ref{eq:g(E)}, is the DOS of graphene in the $x-y$ plane.

\section{Quantum Rate and Density-of-States of Graphene}\label{sec:DOS-graphene}

It must be noted that $g(E)$ of Eq.~\ref{eq:g(E)} corresponds to the DOS $\left( dn/dE \right)$ and can be combined with the result of Eq.~\ref{eq:Cq}. This yields

\begin{equation}
\label{eq:Cq(E)}
	C_q(E) = e^2 g(E) = g_s g_v e^2 \frac{A}{2\pi} \left[ \frac{E}{\left( \hbar c_* \right)^2} \right].
\end{equation}

\noindent This allows for the definition of $C_q$ as a function of $E$. Consequently, Eq.~\ref{eq:Cq(E)} defines a $\tau(E) = 1 / \nu(E) = R_qC_q = G_0/C_q(E)$ that, according to Eq.~\ref{eq:nu-G}, can be stated as

\begin{equation}
\label{eq:tau-q(E)}
	\tau(E) = g_v \frac{A}{\hbar} \left( \frac{E}{c_*^2} \right).
\end{equation}

Given that $|\textbf{k}|_{x-y}/E = d|\textbf{k}|_{x-y}/dE = 1/\hbar c_*$, the function $\tau(E)$ represented in Eq.~\ref{eq:tau-q(E)} can be also expressed as a linear function of $|\textbf{k}|_{x-y}$, that is,

\begin{equation}
\label{eq:tau-q(k)}
	\tau(|\textbf{k}|_{x-y}) = \left( \frac{g_v A }{c_*} \right) |\textbf{k}|_{x-y},
\end{equation}

\noindent where $\tau(|\textbf{k}|_{x-y}) = 1/\nu(|\textbf{k}|_{x-y})$ has a direct proportionality with the particle's momentum $\textbf{p}_{x-y} = \hbar \textbf{k}_{x-y}$.

In summary, the quantum-rate concept $\nu = G_0/C_q = 1/\tau$ defines a characteristic $\tau = R_qC_q$ time constant for the transport of electrons within an electronic structure with a DOS defined as $C_q/e^2$ that, owing to its intrinsic relativistic characteristics, can be stated as a function of both the energy and momentum of the particle's dynamics. In the case of the electronic structure of graphene, the characteristic $\nu$ or $\tau$ as a function of $E$ or \textbf{p} are defined in Eq.~\ref{eq:tau-q(E)} and Eq.~\ref{eq:tau-q(k)}, respectively.

\section{Remarks on the Experimental Analysis of the Electronic Structure}\label{sec:RE-DOS}

Notably, in a photo-emission spectroscopic experiment such as that of angle-resolved photo-emission spectroscopy (ARPES)~\citep{Cheol-2009}, the structure of graphene is revealed by imposing a photonic energetic $h\nu$ perturbation with a threshold frequency $\nu$ that corresponds to the energy level that permits photoelectron emission from $x-y$ plane of graphene. The emitted photoelectron with particular energy and momentum is detected at different angles by using a photodetector and reveals the electronic structure of graphene~\citep{Cheol-2009}. 

In a quantum-rate spectroscopic experiment, a $\tau(E)$ (in agreement with Eq.~\ref{eq:tau-q(E)}) is measured resulting from a time-dependent perturbation of the momentum $\textbf{p} = \hbar \textbf{k}$ (in agreement with Eq.~\ref{eq:tau-q(k)}) of the electron particle in the $x-y$ plane of graphene. This helps reveal the electronic structure due to the phase coherence between the time perturbation of energy $\tau(E)$ and momentum $\tau(\textbf{k})$ with relativistic electrodynamics. This implies that the ratio between the perturbation $E$ and $\textbf{k}$ is coherent such that

\begin{equation}
\label{eq:tau-ratios)}
	\frac{\tau(\textbf{E})}{\tau(\textbf{k})} = \frac{E}{\textbf{p} \cdot \textbf{c}_*} = \frac{\omega}{\textbf{c}_* \cdot \textbf{k}} ,
\end{equation}

\noindent where $\omega = e^2/\hbar C_q$ can be identified as the angular version of $\nu$ defined in Eq.~\ref{eq:nu}. The resting mass $m_0=0$ of Fermions in graphene implies, from a relativistic perspective $E = \sqrt{(\textbf{p} \cdot \textbf{c}_*)^2 + (m_0 c_*^2)^2}$, that $E = \textbf{p} \cdot \textbf{c}_*$.

Accordingly, the goal of the present manuscript is to demonstrate that a time perturbation of $E$ entails a response of $\textbf{k}$ and $vice-versa$. Therefore, by measuring $C_q$, a plot of $E = g_s g_v e^2/C_q$ can be constructed directly as a function of $\textbf{k}$\footnote{Note that the establishment of a relationship between $E$ and $\textbf{k}$ was only possible  because, in graphene, the Dirac point energy state is determined experimentally, and it corresponds to the $K$ spatial point. By invoking spatial symmetry rules and assuming the tight-binding structural model as the reference of the spatial information in the $x-y$ plane of graphene, the $M$ and $\Gamma$ positions were estimated with respect to the referential $K$ spatial Dirac point position. This helped plot the energy \textit{versus} spatial position for the quantum-rate spectroscopy (QRS) method, as shown in Figure~\ref{fig:ES-comparisons}.}. The plot of $E$ \textit{versus} $\textbf{k}$ corresponds, in good quantitative agreement, to the electronic structure computationally calculated through DFT or measured using ARPES.

\section{Experimental and Computational Methods}\label{sec:Exp+Comp}

\subsection{Experimental Methods}\label{sec:Experimental}

SLG on 90-nm SiO$_2$/Si substrate, as provided by Graphenea company, was obtained via chemical vapor deposition. Electric contacts for an electrochemical frequency-response analysis were made on SLG surfaces with sizes of 10 \textit{versus} 10 mm. The contacts consisted of an area of 8 \textit{versus} 2 mm and were fabricated via radio frequency sputtering consisting of a deposition of a first layer of titanium of $\sim$ 10 nm thickness over which a layer of gold of $\sim$ 100 nm thickness was deposited. 

The frequency-response analysis was conducted using a portable potentiostat (PalmSens4) equipped with a frequency-response analyzer. The analysis consisted of a small potential sinusoidal perturbation $\tilde{V} = |V_0| \exp \left(j \omega t \right)$ with a root mean square potential amplitude $|V_0|$ of 10 mV over a potential bias $\bar{V}$ measured concerning the Fermi-level potential of electrode $E_F/e$ (here corresponding to the Dirac point of graphene), where $\omega$ is the angular perturbation frequency and $j = \sqrt{-1}$. The supporting electrolyte of the electrochemical cell was constituted of 1-butyl-3-methylimidazolium tetrafluoroborate [BMIM]BF$_4$ ionic liquid (IL) with purity $\sim 98\%$ (Sigma Aldrich) that acts as a dielectric layer over the graphene plane. The measurements were conducted at steady-state potential $\bar{V}$ with perturbing frequencies ranging from 1 MHz to 3 Hz or at a fixed frequency with potential scanned positively or negatively to the Fermi level potential $E_F/e$.

Notaby, for the frequency-response analysis, an electric current response $I(t) = \bar{I} + \tilde{I}$ was measured as a response to the $V(t) = \bar{V} + \tilde{V}$ time potential perturbation. The ratio between the potential perturbation and current response help perform a transfer-function spectroscopic analysis. For instance, an admittance of $G^*(\omega) = \tilde{I}/\tilde{V} = j\omega C^*(\omega)$ or impedance $Z^*(\omega)= \tilde{V}/\tilde{I} = 1/[j\omega C^*(\omega)]$ complex function can be considered a particular electric transfer-function. From the aforementioned functions, the complex $C^*(\omega)$ capacitive spectra are of particular interest for studying graphene, as discussed in subsequent sections concerning the analysis of Eq.~\ref{eq:Cq-complex} and its related spectra.

For the measurement of $C^*(\omega)$ spectra of SLGs, graphene layers were embedded in an electrochemical cell by using a two-electrode configuration. Here, SLG served as the working electrode, and a platinum mesh served as the counter/reference electrode; for SLGs perturbed in a time-dependent voltage or current mode of measurement, a particular electrochemical cell configuration was employed. This cell configuration employs a graphene layer serving as the working electrode in a particular quantum transistor configuration, wherein the electric current flowing in the channel of the graphene transistor is ``purely’’ capacitive with the series equivalent capacitance dominated by the quantum capacitance. Hence, this electrochemical cell configuration corresponded to using graphene layers as quantum AC transistors (specifically as a diode mode of an AC transistor configuration). 

Therefore, unlike DC-transport where particle--wave currents cannot be investigated, here, the time-dependent transport of the total current $\textbf{j}(\textbf{r})$ is the sum of the displacement current $\epsilon \left( \delta \textbf{E} / \delta t \right)$ and particle $\textbf{j}_p(\textbf{r})$. This leads to

\begin{equation}
\label{eq:j-currents}
	\textbf{j} \left( \textbf{r} \right) = \epsilon \left( \frac{\delta \textbf{E}}{\delta t} \right) + \textbf{j}_p \left( \textbf{r} \right),
\end{equation}

\noindent where $\textbf{r}$ corresponds to the position vector of the particle, $\epsilon$ (considered space- and time-invariant for simplicity) is the dielectric constant, and \textbf{E} is the dielectric field. The total current of Eq.~\ref{eq:j-currents} must be conserved, such that

\begin{equation}
\label{eq:div-j}
	\nabla \cdot \textbf{j} \left( \textbf{r} \right) = 0.
\end{equation}

Eq.~\ref{eq:div-j} states that along a line tangential to the current vector \textbf{j}, the length of this vector is invariant. Therefore, similarly to the conservation law of energy that permits the transformation of kinetic into potential energy, the particle current is transformed into displacement current and \textit{vice-versa}. Notably, while the particle current exists only inside the electric conductors that are to be examined, the displacement current is not limited to the conductor and can be detected through an external probe---this aspect is crucial for the analysis in this study. 

In other words, for a time-dependent electric field perturbation, a displacement current can be measured by a contact probe. If this measured displacement current results from a particle perturbation response inside the conductor (here, graphene), the particles are detected as a result of the displacement current response to the perturbation. This analysis is a straightforward consequence of particle--wave duality associated with relativistic characteristics within Eq.~\ref{eq:nu} and Eq.~\ref{eq:Planck-Einstein} and of phase coherence between the perturbation and response.

Accordingly, the use of an electrochemical AC transistor configuration enables the measurement of the properties of the channel of the transistor by short-circuiting the drain and source contact; the capacitive property of the channel results from a field effect obtained by the AC response of the channel to the external field-perturbation. This configuration of measurement was successfully tested and previously reported as a suitable way of obtaining, for instance, the conductive and capacitive V-shapes of graphene~\citep{Lopes-2021-1}.

Here, the analysis of $C^*(\omega)$, which theoretically correlates with Eq.~\ref{eq:tau-q(E)} and Eq.~\ref{eq:tau-q(k)}, helps analyze the electronic structure of graphene, as shown later. Complex capacitance spectra can be constructed either from the raw admittance or impedance data. For instance, using impedance data as an example, capacitive spectra can be obtained by converting the complex impedance $Z^*(\omega)$ function into a complex capacitance $C^* (\omega)$ using the relationship $C^* (\omega)= 1/[j\omega Z^*(\omega)]$. The real and imaginary components of $C^* (\omega)$ are $C^{'}=Z{''}/\omega|Z|^2$ and $C{''}=Z{'}/\omega|Z|^2$, respectively, where $|Z|=\sqrt{Z{'}^2 + Z{''}^2}$ is the modulus of $Z^*(\omega)$.

\subsection{Computational Methods}\label{sec:Computational}

The electronic structure of SLG containing 180 atoms was calculated in a vacuum using first-principles calculations within a DFT framework~\citep{Hohenberg-1964}, as implemented in the SIESTA code by employing a generalized gradient approximation (GGA) in its PBE formulation for the exchange and correlation potential. It used a finite real-space grid corresponding to a 200-Ry plane-wave cut-off for the integrals in real space.

The Kohm--Sham wave functions~\citep{Kohn-1965} for the valence electrons were expanded in a double-zeta-polarized (DZP) basis set of soft-confined finite-support numerical atomic orbitals~\citep{Junquera-DZP}. The band-structure calculations were performed using SIESTA code~\citep{Sanz-Navarro-2011}, sampling the Brillouin zone with k-points along the $\Gamma \rightarrow M \rightarrow K \rightarrow \Gamma$ path, as detailed in the Supporting Information document.

\section{Results and Discussions}\label{sec:Results}

\begin{figure*}[hb]
\centering
\includegraphics[width=14cm]{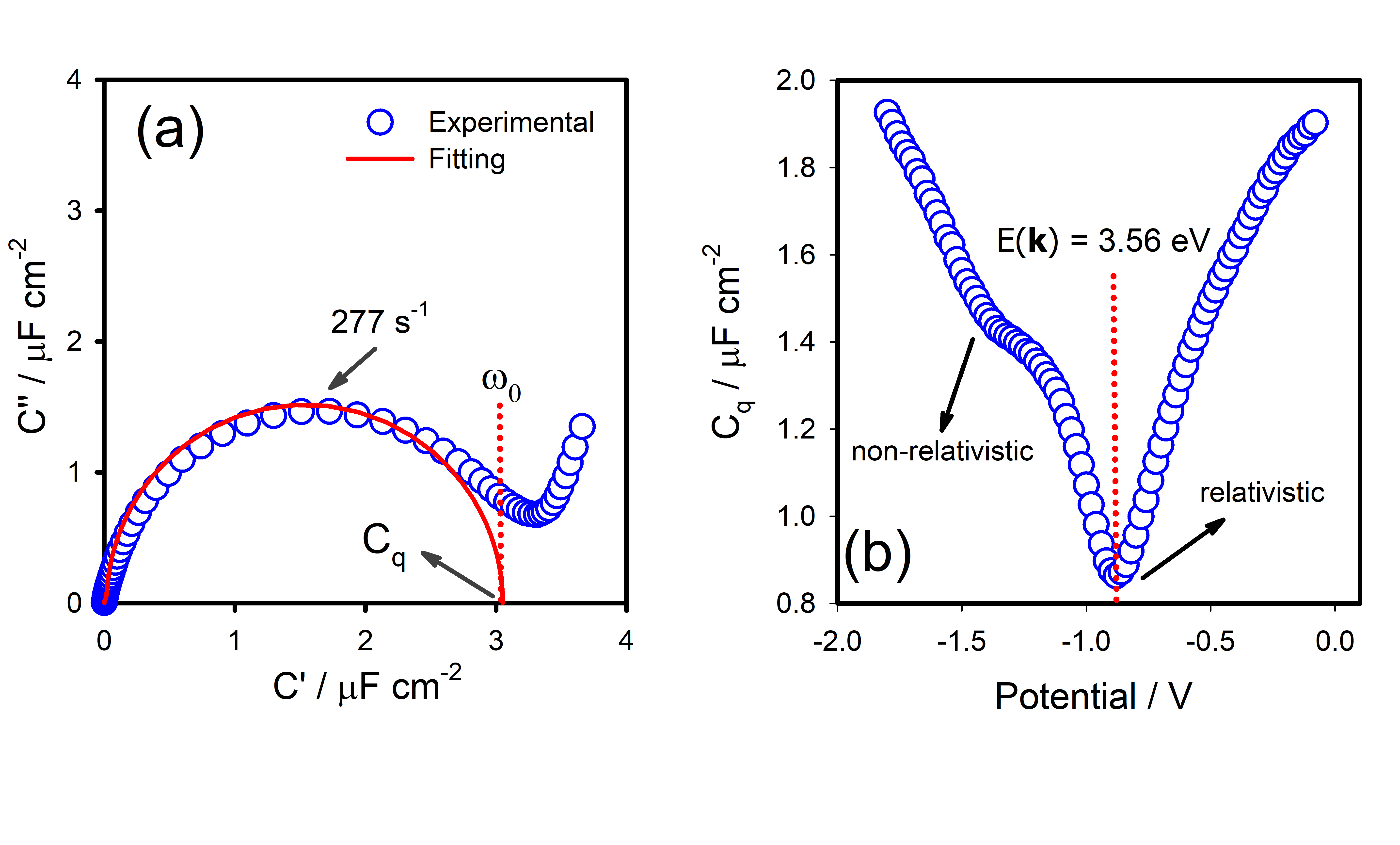}
\caption{(a) Nyquist capacitive diagram of a SLG graphene. A resonance frequency of 277 s$^{-1}$ whereas a $\omega_0/2\pi \sim$ 13 s$^{-1}$ as the low-frequency limit where $C^0_q$ can be obtained according to Eq.~\ref{eq:Cq-complex}. A fitting to Eq.~\ref{eq:Cq-complex} considering an additional series resistance contribution in that $R_q$ is obtained as the total series resistance $R_{s,t}$, and this corresponds to the theoretical limiting value of $R_q \sim$ 12.9 k$\Omega$, as discussed section~\ref{sec:RC}. (b) depicts the capacitive V-shape obtained at $\Omega_0$ limiting frequency.}
\label{fig:Nyquist+CVshape}
\end{figure*}

As predicted by Eq.~\ref{eq:Cq(E)}, the electronic structure of graphene is directly proportional to $C_q(E)$. Therefore, in agreement with the quantum-rate theory, as stated in Eq.~\ref{eq:nu}, the measurement of a complex capacitive spectra $C^*(\omega)$ of graphene must correspond to the response of a quantum resistive-capacitive (RC) such as that in~\citep{Bueno-book-2018}

\begin{equation}
 \label{eq:Cq-complex}
	C_q^*(\omega) = \frac{C^0_q}{1 + j\omega \tau} \sim C^0_q \left( 1 - j\omega \tau \right) + O(\omega^2),
\end{equation}

\noindent as demonstrated in previous works~\citep{Lopes-2021-1, Lopes-2021-2}. Notably, a $C_q^*(\omega)$ spectrum of graphene can be measured by a direct frequency-response analysis. 

The $C_q^*(\omega)$ spectrum corresponds to the capacitive frequency response of a quantum RC equivalent circuit that can be expressed, according to the quantum-rate theory, solely as a response of series $R_q$ and $C_q$ quantum circuit elements, thus providing a characteristic frequency $\omega_c = 1/(R_q C_q)$ with a characteristic time of $\tau = 2\pi R_q C_q$. 

In other words, in agreement with the relativistic quantum mechanics predicted by the quantum-rate theory, Eq.~\ref{eq:nu}, the quantum electrodynamics of graphene can be studied in detail by using a simple electrochemical experimental setup. As an example, it is herein demonstrated that such type of experiments permit the development of an \textit{in-situ} spectroscopic method of measuring the electronic structure of graphene, although the method is not limited to revealing the electronic structure of graphene. The electronic structure of molecular-scale organic~\citep{Bueno-2023-3} and inorganic~\citep{Pinzon-2021} films as well as quantum dots~\citep{Pinzon-2023} can also be revealed using quantum-rate spectroscopic method.

\subsection{Quantum RC Circuit and Quantum Capacitance of Graphene}\label{sec:C-spectrum}

Eq.~\ref{eq:Cq-complex} permits the evaluation of the quantum-rate characteristic response of graphene in a critical quantum electrodynamic setting. This corresponds to a particular frequency $\omega_0$ (see Figure~\ref{fig:Nyquist+CVshape}), where the electronic structure is measured (or responds to a perturbation) in a quasi-equilibrium condition. Mathematically, this corresponds to $\omega_0$ tending to null, wherein a $C^0_q$ value can be evaluated. 

From a theoretical physical analysis, $C^0_q$ measured at $\omega_0$ corresponds to a frequency with minimal influence of the imaginary component of $C^*(\omega)$ spectra. In this case, there is a dominant response of the real component of $C^*(\omega)$ that corresponds to $C^*(\omega)$ responding to a limiting perturbation frequency $\omega_0$, as indicated in Figure~\ref{fig:Nyquist+CVshape}\textit{a}.

\begin{figure}[hb]
\centering
\includegraphics[width=7cm]{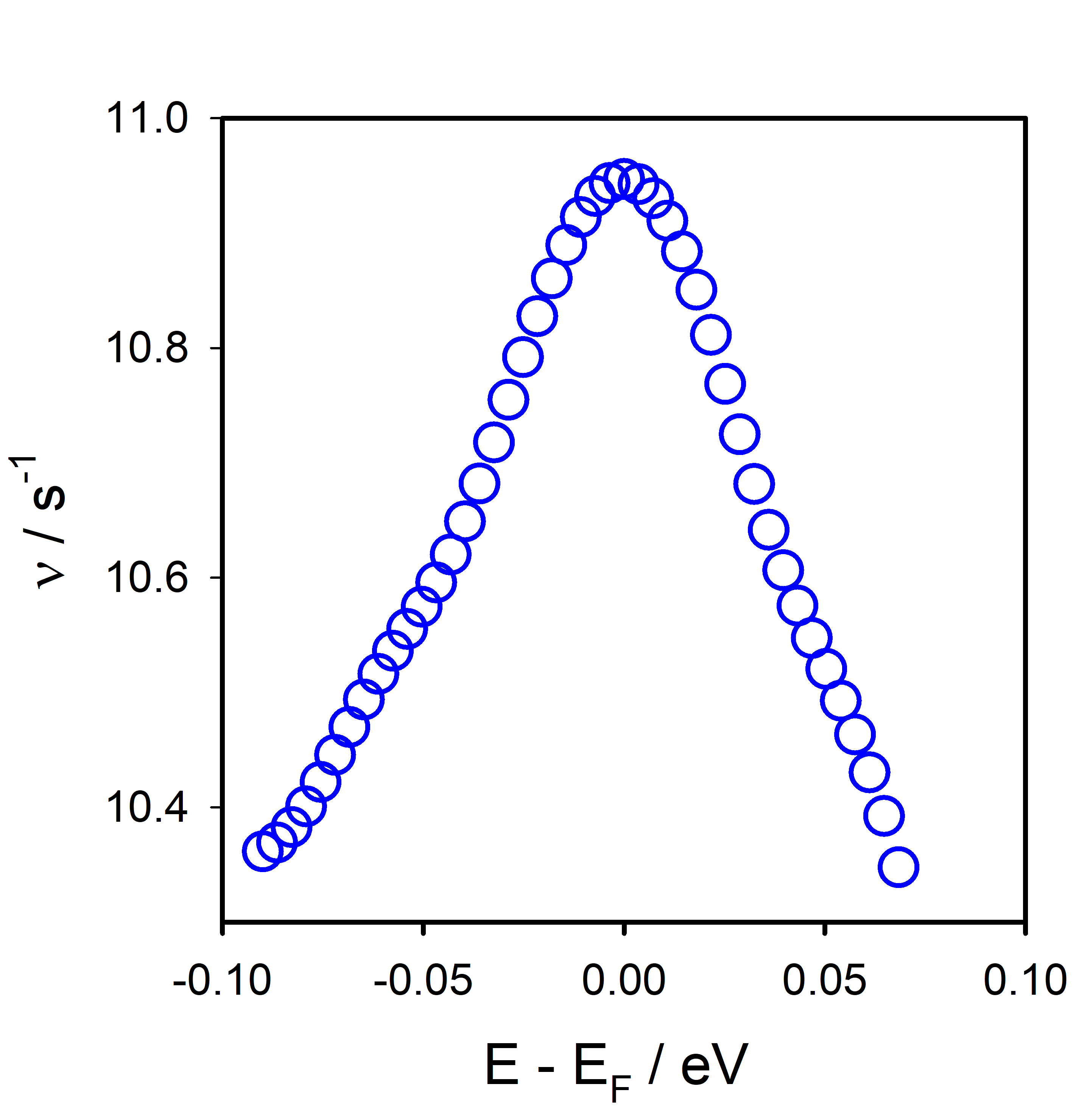}
\caption{$\nu = G_0/C_q$ measured for an SLG around the Dirac-point. As theoretically expected, this $\nu$ is the maximum at the Dirac point owing to $C_q$, and the conductance is maximum. This is the theoretically expected result as, in contraction to the common classical sense for an RC circuit, for a quantum RC circuit, the resistance is the minimum (corresponding to a maximum conductance) for a maximum relaxed capacitance.}
\label{fig:Qrate-EF}
\end{figure}

Nonetheless, from a quantum mechanical RC perspective, at this limiting frequency $\omega_0$, the imaginary component of $C_q^* = jC'' + C'$, correlated to $G(\omega_0) = \omega_0C''$ is not zero, as expected in a classical analysis of a series RC circuit, corresponding to a null electric current circulating in the resistor. According to the quantum-rate theory, at the limit of $\omega_0$, the quantum RC circuit analysis corresponds to the minimum quantum resistance of the circuit with a non-zero current in the conductor. In other words, a quantized displacement current exists in the resistor, and this quantum resistive current is measured at $\omega C''$ and is associated with the value of $C_q^0$. 

This electric current is non-zero and is likely associated with multiples of conductance $e^2/h$; it enables the determination of a quantum limit of perturbation to the energy of the system $E_0 = e^2/C_q^0$. Owing to the resistance being quantized and difference of zero (with multiples of $n$ states), it permits coherent dynamics between the perturbation of potential (or energy) and current (or quantum states) response for an equilibrium charge condition $C_q^0 = ne/V_0$. This equilibrium charge relaxation between the perturbation of the probe and response of the quantum RC system is an important physical characteristic making quantum-rate spectroscopy useful. As expected for classical RC circuit analysis, at low frequencies, the current in the resistor is expected to be null or dissipative and not coherent with the potential perturbation. This is not the case in a quantum RC circuit analysis. 

Therefore, at the Dirac point level of energy, the resistance examined at the limit of $\omega_0$ must correspond to the resistance quantum $R_q = 1/G_0 \sim$ 12.9 k$\Omega$. This is owing to $L = \lambda$, that is, a situation corresponding to perturbing the single $n$ quantum state mode of the quantum channels of graphene (with lengths $L_x$ and $L_y$, as noted in Figure~\ref{fig:E-k} in the plane) at the Dirac point.

\begin{figure*}[h]
\centering
\includegraphics[width=16cm]{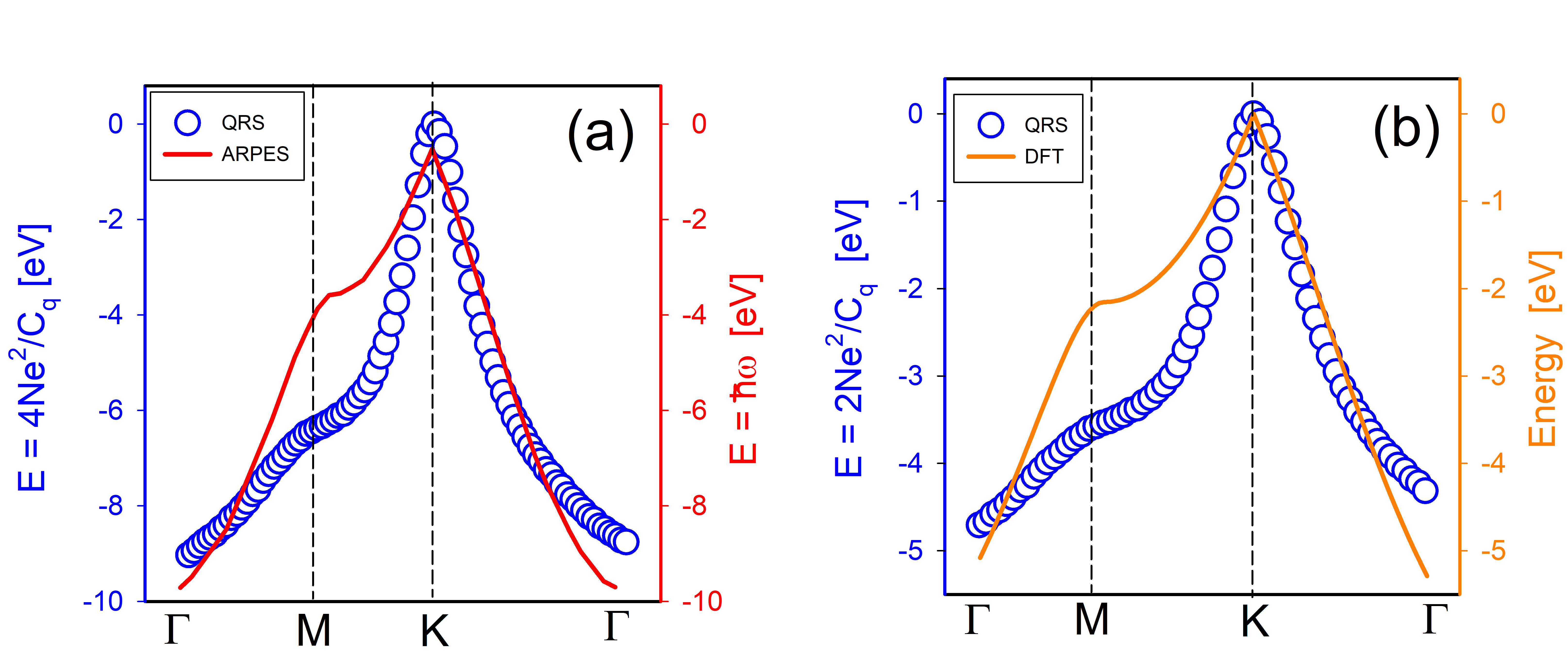}
\caption{(a) Blue dots indicate the electronic structure of an SLG graphene as measured by QRS; the red curve indicate the values obtained via ARPES. (b) Electronic structure obtained via QRS (blue dots) compared with that calculated through DFT (in orange). In both (a) and (b), there is an excellent agreement between the QRS method and ARPES or DFT. The differences observed mainly around the $M$ position are attributed to the [BMIM]BF$_4$ electrolyte, as discussed in section~\ref{sec:QRS-spectrum}.}
\label{fig:ES-comparisons}
\end{figure*}

The mean $G$ value measured over three different samples (see SI document for more information) was obtained as $G (\omega) = \omega_0 C'' \sim$ 19 $\pm$ 0.9 mS that multiplied by $g_s$ and $g_v$, as required for the consideration of energy degeneracies, led to $\sim$ 76.1 $\pm$ 1.8 mS. This mean experimental value corresponds to 2\% of deviation of the theoretical expected value of $G_0 = g_s e^2/h \sim$ 77.5 mS.

Additionally, the experimental value from the analysis of the whole $C^*(\omega)$ spectra obtained at the Fermi level (for three different SLG samples), as shown in Figure~\ref{fig:Nyquist+CVshape}, corresponded to $R_q = 1/G_0 \sim$ 12.9 $\pm$ 1.6 k$\Omega$, in excellent agreement with the theory that predicts a value of $\sim$ 12.9 k$\Omega$. This resistance quantum limit value was obtained from the fitting of the spectra to Eq.~\ref{eq:Cq-complex}, and $R_q$ in Eq.~\ref{eq:Cq-complex} was considered as the total series resistance of the circuit encompassing the resistance of the contacts; see SI document for more details. 

The aforementioned method of calculating $R_q$ is similar to that used successfully in the analysis of Faradaic quantum conductance of redox-active monolayers~\citep{Sanchez-2022-2, Bueno-2023-3}. As observed in the electrochemical experiments of redox-active monolayers~\citep{Sanchez-2022-2}, in the analysis of graphene, the total current (indicated in Eq.~\ref{eq:j-currents}) is experimentally accounted for, and not only the particle's current $\textbf{j}_p$ exclusively associated to the graphene. This is in agreement with Eq.~\ref{eq:div-j}, and the quantum circuit cannot be treated separately from the external circuit.

In the case of graphene, it was considered that $g_s$ and $g_v$ degeneracies, such as the experimental $\omega_{c,e} = 1/(R_{s,t}C_q)$, equate to the theoretical characteristic frequency of $\omega_c = g_s g_v e^2 / \hbar C_q$, where $R_{s,t}$ is the total series resistance, and $R_{s,t}$ equates to the resistance quantum including the contact resistance. By equating $\omega_{c,e}$ to $\omega_c$ a value of $R_q = \left[2\pi (g_s g_v) R_{s,t} \right] \sim$ 12.9 $\pm$ 1.6 k$\Omega$ is obtained. Notably, $g_s$ and $g_v$ were obtained from theoretical analyses, whereas the measured resistance quantum effectively measured for graphene at the Dirac point was $h/e^2$, as reported previously~\citep{Lopes-2021-1}.

As demonstrated in previous study, conductive and capacitive V-shapes for graphene can be obtained using time-dependent analysis~\citep{Lopes-2021-1}. These V-shapes were obtained at a minimal perturbation of the electronic structure, that is, by measuring $C_q^0$ at $\omega_0$ as a function of the energy $E = -eV$ of the electrode from where electrons and holes can be injected/ejected in an equilibrium charge condition of quantum RC relaxation. The smallest value of $C_q^0(E)$ is obtained at the Dirac point; however, it cannot be zero in accordance with Eq.~\ref{eq:nu}; it would correspond to a zero Fermi-energy, which would be physically incorrect, as better argued in a previous work~\citep{Bueno-2022}. 

The minimal of energy corresponds to $\Delta E(\textbf{k}) = E - E_F = 0$, and hence $E(\textbf{k}) = E_F(\textbf{k}) = -3.56$ eV to the vacuum energy reference level; this shows good agreement with the theoretical value of $-$3.45 eV calculated in this study by using DFT computational method. The approximately 10 meV differences between $-$3.45 and $-$3.56 eV can be attributed to the experimental $-$3.56 eV value obtained for an SLG measured in ionic-liquid ([BMIM]BF$_4$) electrolyte environment wherein $-$3.45 eV was calculated via DFT for an SLG placed in vacuum.

In other words, as detailed in a previous work~\citep{Bueno-2022}, $C_q^0$ cannot be zero as this would imply an infinite value of $E_F$. Although the energy at the Fermi level or Dirac point must be maximal with a minimal $C_q^0$, as shown in the capacitive V-shape of Figure~\ref{fig:Nyquist+CVshape}\textit{b}, this minimal value of $C_q^0$ cannot be zero, as suggested by some researchers~\citep{Tao-2009}. This inference for a zero capacitance at the Dirac point was based on an analysis of the absence of a net charge carrier density at the Dirac point with no polarization of the graphene; thus, no net charge carriers are present to charge a capacitance~\cite{Tao-2009}. The net number of charge carriers is zero at the Dirac point; however, this implies an ambivalent displacement electric current according to Eq.~\ref{eq:j-currents} that permits examination of the particle dynamics even in the absence of a net DC current. Furthermore, the time-dependent quantum analysis of this problem indicates that the quantum mode $n$ can correspond to quantum mode of the ground state where there is no net current. This analysis is corroborated by the value of $e^2/h$ determined by the quantum of conductance at the Dirac point, corresponding to the ground state fermionic energy $e^2/C_q^0$ with the minimum value but different from a null capacitance, as suggested by a prior analysis~\cite{Tao-2009} from the charge density perspective to interpret the quantum capacitance of graphene at the Dirac point.

According to the implications of the aforementioned $C_q^0$, although there are no net charge carriers at the Dirac point, the value of $C_q^0$ cannot be zero owing to statistical and quantum mechanical reasons. Accordingly, the non-zeroed minimal value for $C^0_q$ obtained experimentally implying null net charge carriers cannot be null from a quantum mechanical perspective, wherein it would be related to a limiting value of resistance governed by the quantum limit of $R_q$ around $E_F = e^2/C_q^0$ with a non-zero minimum $C^0_q$ owing to a quantum resistance limit. The minimum $R_q$ and $C_q$ predicted values at the Dirac point conduct to the maximum rate $\nu \propto e^2e/hC_q^0 = E_F/h$, as shown in Figure~\ref{fig:Qrate-EF}. Notably, albeit the minimum capacitance $\sim$ 0.83 $\mu$F cm$^{-2}$ obtained from a V-shaped capacitive measurement at $\omega_0$, as observed in Figure~\ref{fig:Nyquist+CVshape}\textit{b}, is in good agreement with the expected statistical mechanical theoretical value of $C_q^0$ of 0.8 $\mu$F cm$^{-2}$; this value can vary according to the electrolyte-imposed external potential and can be lower than 0.8 $\mu$F cm$^{-2}$ in certain circumstances but not zero.

The next section demonstrates how the electronic structure of graphene is obtained using the quantum-rate spectroscopic concept.

\section{Quantum Rate Spectroscopy of Graphene}\label{sec:QRS-spectrum}

It must be noted that Eq.~\ref{eq:tau-ratios)} is equivalent to $E~\textbf{k} = \omega~\textbf{p}$, being a different expression of De Broglie or Einstein--Planck relationships. This equation is also equivalent to $E \tau$ = $|\textbf{p}| \lambda$, implying that there is a phase coherence between the perturbation and the response to the perturbation, as expected from the quantum coherent RC circuit analysis, as discussed in previous sections.

To resolve any questions related to the principles of a spectroscopic method based on the quantum-rate concept, defined in Eq.~\ref{eq:nu}, let us demonstrate the correspondence between the duality particle $E = e^2/C_q$ (or $p$) and $\textbf{k}$ ($\omega$) wave of the electron whenever there is a time-dependent perturbation on $C_q$ that corresponds to a coherent response of $\textbf{k}$ and \textit{vice-versa}.

Hence, let us consider Eq.~\ref{eq:Cq}, which can be rewritten as

\begin{equation}
 \label{eq:dE/dn}
	\left( \frac{dE}{dn} \right) = \frac{e^2}{C_q},
\end{equation}

\noindent where $e^2/C_q = \hbar \textbf{c}_* \cdot \textbf{k} = E$ is the energy associated with $C_q$, according to Eq.~\ref{eq:Planck-Einstein} and premises of the quantum-rate theory. Evidently, $dE = \hbar \textbf{c}_* \cdot \textbf{k} dn$ and a time perturbation in $E$ corresponds to a time perturbation of $n = q/e = L/\lambda$ such that

\begin{equation}
 \label{eq:dE/dt}
  \left( \frac{dE}{dt} \right) = \hbar \textbf{c}_* \cdot \textbf{k} \left( \frac{dn}{dt} \right),
\end{equation}

\noindent where $E$ and $n$ are the only time-dependent variables; $dE/dt = E(t)$ and $dn/dt = n(t)$. Therefore, an energy perturbation $E(t)$ of a quantum RC system as a function of time corresponds to a coherent response of quantum RC states $n(t)$ as a function of time. These time-dependent and related functions can be stated as $E(t) = \bar{E} + \tilde{E}$ and $n(t) = \bar{n} + \tilde{n}$, respectively, where $\tilde{E} = E_0 \exp \left(j\omega t \right)$ and $\tilde{n} = n_0 \exp \left(j\omega t - \phi \right)$ are the oscillatory perturbation of $E$ and the corresponding oscillatory response of $n$ to a defined $E$ perturbation. 

$E_0$ and $n_0$ are the amplitudes of the energy perturbation of the quantum RC state response, respectively, and $\phi$ is the coherent phase difference between energy perturbation and quantum RC states response. $\bar{E}$ and $\bar{n}$ denote steady-state levels of the system where the oscillatory perturbation is performed. In the present situation of measurement, conducted in graphene, steady-state energy levels $\bar{E} = -eV + E_F$ correspond to the positive or negative bias voltage $V$ applied to the Dirac point $E_F$.

The aforedescribed analysis shows that $E(t) = E_0 \exp \left( j\omega t \right)$ and $n(t) = n_0 \exp \left( j\omega t - \phi \right)$ allows defining a complex $E^{*}(\omega)$ energy state function such as

\begin{equation}
 \label{eq:dE/dn-complex}
	E^*(\omega) = \left[ \frac{dE(t)}{dn(t)} \right] = \left( \frac{E_0}{n_0} \right) \exp \left(j\phi \right),
\end{equation}

\noindent wherein the real component of $E^*(\omega)$ is identified as

\begin{equation}
 \label{eq:dE/dn-real}
	\Re \left[ E^*(\omega) \right] =  \left( \frac{E_0}{n_0} \right) \Re \left[ \exp \left(j\phi \right) \right] = \hbar \textbf{c}_* \cdot \textbf{k}.
\end{equation}

As discussed previously, if $\omega$, in Eq.~\ref{eq:dE/dn-real}, is considered $\omega_0$, there is a limit for the energy $E_0 = e^2/C_q^0$ that corresponds to a limit of Eq.~\ref{eq:Cq-complex} in which $e^2/n_0C_q^0 = \hbar \textbf{c}_* \cdot \textbf{k}$, \textit{quod erat demonstrandum}. This previous demonstration implies a correspondence between $e^2/C_q^0$ and $\hbar \textbf{c}_* \cdot \textbf{k}$ that allows for the measurement of the electronic structure of graphene \textit{in-situ} by using an electric time-dependent method.

\begin{figure*}[h]
\centering
\includegraphics[width=16cm]{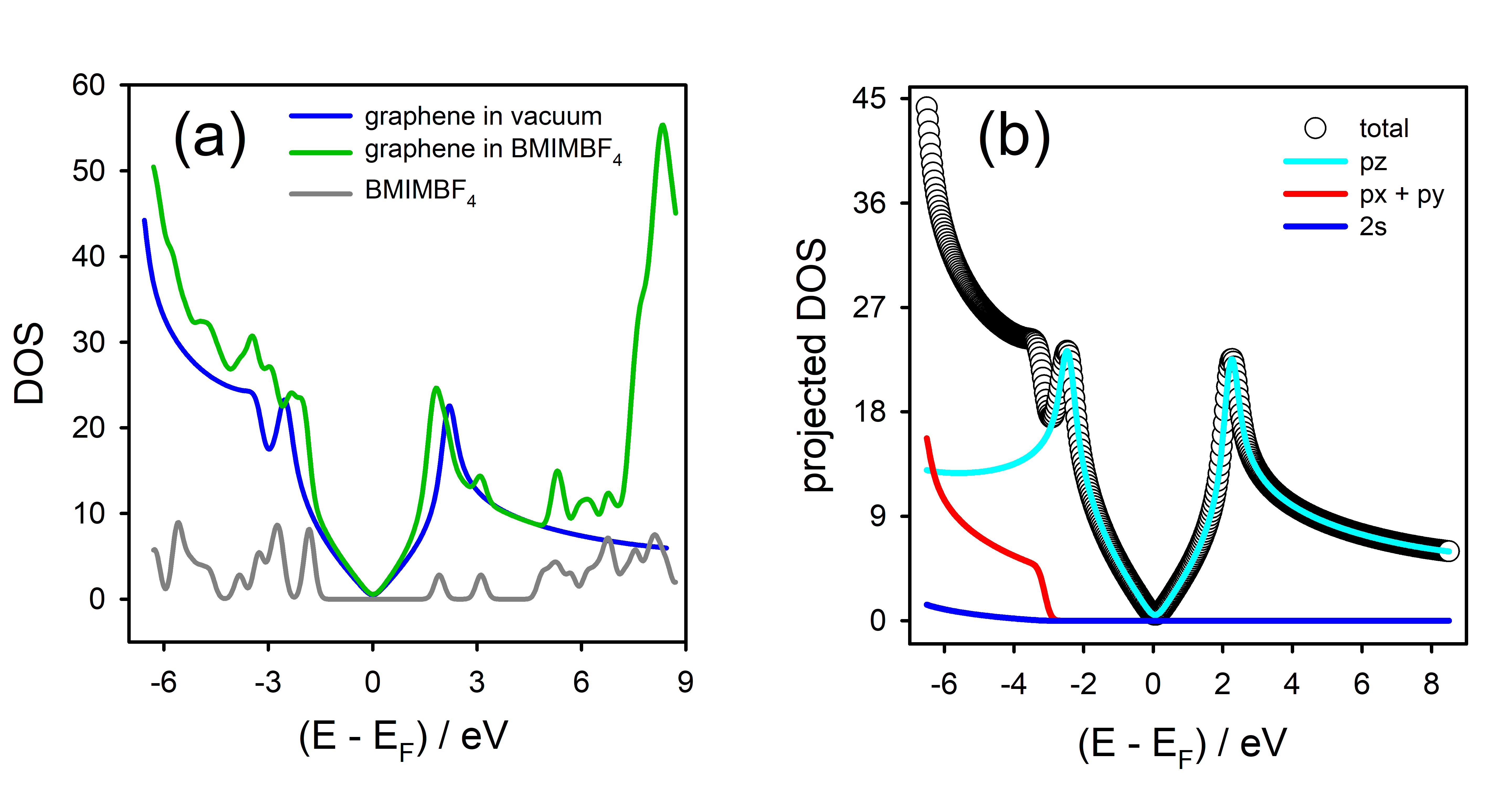}
\caption{(a) Analysis results of the individual contribution of [BMIM]BF$_4$ to the DOS of graphene in contact with this electrolyte. Evidently, the presence of [BMIM]BF$_4$ affects the electronic structure of graphene measured in the presence of [BMIM]BF$_4$ and was identified as the main difference between the electronic structure of graphene measured in vacuum by ARPES and that measured by QRS in the presence of [BMIM]BF$_4$. (b) Projected DOS of graphene in a vacuum calculated via DFT with different individual contributions of the orbitals to the total response indicated in black dots.}
\label{fig:DOS-DFT}
\end{figure*}

The capacitive V-shape depicted in Figure~\ref{fig:Nyquist+CVshape}\textit{b}, obtained at $\omega_0$, is the basis to calculate the electronic structure of graphene, following physical concepts discussed in Eqs.~\ref{eq:tau-q(E)} to Eq.~\ref{eq:tau-ratios)}. Accordingly, using the capacitive V-shape of Figure~\ref{fig:Nyquist+CVshape}\textit{b} and the relationship $e^2/n_0C_q^0 = \hbar \textbf{c}_* \cdot \textbf{k}$, the electronic structure of graphene is revealed in Figure~\ref{fig:ES-comparisons}. The represented structure is associated with 2D electronic structure of graphene, i.e., obtained considering the $x-y$ plane and ignoring the thickness. Nonetheless, the experimental potential electric perturbation (with a very low excitation energy) was conducted, as discussed in previous section, in the thickness and consequently in the z-direction perpendicular to the $x-y$ plane. Note that whereas the ordinate axis was obtained experimentally in Figure~\ref{fig:ES-comparisons}, the abscissa axis was estimated using 2D structural tight-binding model considering the symmetry spatial structure of graphene with respect to the Dirac point which can be measured experimentally using the QRS method. Albeit the calculation of abscissa axis is an approximation, it provides, particularly for graphene, a very good correlation with other methods, as noted by comparing the results of QRS to DFT and ARPES methods. 

In Figure~\ref{fig:ES-comparisons}\textit{a}, the electronic structure of SLG measured by quantum-rate spectroscopic (QRS) method in electrolyte and room-temperature environment is compared with that obtained by ARPES~\citep{Cheol-2009} in vacuum and low-temperature. In Figure~\ref{fig:ES-comparisons}\textit{b}, the electronic structure measured by the QRS method is compared with that calculated by DFT computational method. As noted, the measured electronic structure by QRS is in good agreement with those obtained theoretically using DFT and experimentally measured using ARPES. In addition, the absolute agreement between the scale of energy $E_0 = e^2/C_q^0$ obtained by DFT and QRS methods depends on the consideration of $g_v$. As the DFT method does not consider $g_v$ in the calculation of the electronic structure, a comparison of the QRS method with that obtained by DFT requires QRS to not consider $g_v$ as well, and so there is a very good agreement between both methods.

The differences between QRS and ARPES/DFT spectra observed in the $M$ position were confirmed via DFT calculations to be attributable to the presence of an electrolyte ([BMIM]BF$_4$) environment over SLG, as noted in Figure~\ref{fig:DOS-DFT}\textit{a}. This figure depicts the comparison between the DOS of graphene placed in the vacuum environment, a graphene in [BMIM]BF$_4$ environment, and the DOS of [BMIM]BF$_4$ isolated from graphene. Evidently, at energies around $(E - E_F) \sim -3.56$ eV + 1.5 eV $\sim -2$ eV, there is a strong contribution of the DOS of the [BMIM]BF$_4$ to that of graphene that affect the total DOS of graphene measured in [BMIM]BF$_4$, as indicated in the figure.

Finally, Figure~\ref{fig:DOS-DFT}\textit{b} depicts the project DOS within the individual contributions of the graphene's orbitals separately. This figure shows that the $pz$ component, which corresponds to an orbital direction perpendicular to that of $x-y$ plane of graphene, is the most affected by the adsorption of [BMIM]BF$_4$ over the $x-y$ plane, as indicated in red, which explains the difference in the electronic structure around $M$ position, as depicted in Figure~\ref{fig:ES-comparisons}.

In summary, QRS facilitates a highly accurate measurement of the electronic structure of graphene in a \textit{in-situ} manner at room-temperature electrolyte environment conditions. This is a notable advantage afforded by this spectroscopic method over others that necessarily involve expensive equipment and time-consuming procedures.

\section{Final Remarks and Conclusions}
\label{sec:conclusions}

The principles of the electrochemical quantum-rate spectroscopy, based on a quantum-rate principle that permits to study relativistic electrodynamics of graphene with great precision, are demonstrated. Although the focus of the present study was on describing a spectroscopic method of measuring the electronic structure of graphene compared with theoretical calculations (DFT) and other experimental (ARPES) methods, we could also demonstrate that graphene has a quantum conductance $e^2/g_v h \sim 19.3$ k$\Omega$ (see SI document for more details) at the Dirac point that is in phase coherence with its equilibrium DOS state (i.e., $(dn/dE) = C_q^0/e^2$), facilitating the establishment of a relationship: $(\hbar \omega_c) (dn/dE) = n_0$. 

The phase coherence analysis of the quantum RC characteristics of graphene, measured at an equilibrium charge dynamics, allowed for measurement of the characteristic energy $E_c = \omega_c \hbar$ of the graphene through a time-dependent perturbation of $n_0$ or $E_0$ as a function of the energy state $E = -eV + E_F = \hbar c_* \cdot \textbf{k}$ of the probe (electrode). 

The plot of $E_c$ as a function of $E \sim eV$ or $\textbf{k}_{x-y}$ presents the essence of the introduced quantum-rate spectroscopic methodology for graphene and enables the \textit{in-situ} measurement of the electronic structure of Weyl semi-metal structures. In other words, although the QRS method is not restricted to graphene, in the present study, the meaning of $C_q$ was defined specifically for graphene in Eq.~\ref{eq:Cq(E)}.

Nonetheless, to study the quantum electrodynamics of other 2D structures at different conditions, the theoretical DOS $\sim C_q/e^2$ of the structure need not be previously known, as the purpose of QRS is to be a simple experimental method for measuring the DOS $\sim C_q/e^2$ of different Weyl semi-metal structures. From an experimental perspective, the presence of an electrolyte is primordial (as detailed in a prior work~\citep{Bueno-2023-3}), because it allows for the equivalent capacitance of the interface $C_\mu$ to be expressed as a function of $C_q$ only---for graphene, it would be written as $1/C_\mu = g_v/C_q$.

\section{Credit author statement}

Laís Cristine Lopes: Data curation, data acquisition and editing, review final draft.

Edgar Pinzón: Data curation and editing, picture-writing, review final draft.

Gabriela Dias-da-Silva: Computational calculation, picture-writing, review final draft.

Gustavo Troinano Feliciano: Computational calculation, 

Paulo R. Bueno: Conceptualization, theory and method development, data curation, writing original and final draft, picture-writing and editing, funding acquisition.

\section{Acknowledgments}

The authors are grateful to Sao Paulo State Research Foundation (FAPESP) for grants 2017/24839-0 and National Council for Scientific and Technological Development (CNPq).

% Use the style of numbering in square brackets.
% If nothing is used, default style will be taken.
%\begin{enumerate}[a)]
%\item 
%\item 
%\item 
%\end{enumerate}  

%\begin{itemize}
%\item 
%\item 
%\item 
%\end{itemize}  

%\begin{description}
%\item[]
%\item[] 
%\item[] 
%\end{description}

% Uncomment and use as the case may be
%\begin{theorem} 
%\end{theorem}

% Uncomment and use as the case may be
%\begin{lemma} 
%\end{lemma}


\printcredits

\bibliographystyle{cas-model2-names}

\bibliography{biblio}

\end{document}